\documentclass[prb,aps,twocolumn,showpacs,nofootinbib,superscriptaddress]{revtex4}

\usepackage{graphicx}
\usepackage{stmaryrd}
\usepackage{rotating}
\usepackage{amsmath}
\usepackage{amsfonts}
\usepackage{amssymb}
\usepackage[countmax]{subfloat}
\usepackage{dcolumn} 
\usepackage{bm} 
\usepackage{color}

\usepackage{float}
\usepackage{latexsym}

\begin{document}

\title{Effect of Quantum Tunneling on Spin Hall Magnetoresistance}


\author{Seulgi Ok} 

\affiliation{Institut f\"{u}r Theoretische Physik, ETH-Z\"urich, CH-8093 Z\"urich, Switzerland}

\author{Wei Chen} 

\affiliation{Institut f\"{u}r Theoretische Physik, ETH-Z\"urich, CH-8093 Z\"urich, Switzerland}

\author{Manfred Sigrist} 

\affiliation{Institut f\"{u}r Theoretische Physik, ETH-Z\"urich, CH-8093 Z\"urich, Switzerland}

\author{Dirk Manske} 

\affiliation{Max-Planck-Institut f$\ddot{u}$r Festk$\ddot{o}$rperforschung, Heisenbergstrasse 1, D-70569 Stuttgart, Germany}

\date{\rm\today}

\begin{abstract}

{We present a formalism that simultaneously incorporates the effect of quantum tunneling and spin diffusion on spin Hall magnetoresistance observed in normal metal/ferromagnetic insulator bilayers (such as Pt/Y$_{3}$Fe$_{5}$O$_{12}$) and normal metal/ferromagnetic metal bilayers (such as Pt/Co), in which the angle of magnetization influences the magnetoresistance of the normal metal. In the normal metal side the spin diffusion is known to affect the landscape of the spin accumulation caused by spin Hall effect and subsequently the magnetoresistance, while on the ferromagnet side the quantum tunneling effect is detrimental to the interface spin current which also affects the spin accumulation. The influence of generic material properties such as spin diffusion length, layer thickness, interface coupling, and insulating gap can be quantified in a unified manner, and experiments that reveal the quantum feature of the magnetoresistance are suggested.
 }

\end{abstract}






\maketitle

\section{introduction}

The electrical control of magnetization dynamics has been a central issue in the field of spintronics\cite{Wolf01,Zutic04}, owing to its possible applications in magnetic memory devices with low power consumption. A particularly promising mechanism for the electrical control is to utilize the spin Hall effect\cite{Dyakonov71,Hirsch99,Sinova04,Sinova15} (SHE) in a normal metal (NM), such as Pt or Ta, to convert an electric current into a spin current, and subsequently to magnetization dynamics in an adjacent magnet via mechanisms such as spin-transfer torque\cite{Berger96,Slonczewski96} (STT). In reverse, the inverse spin Hall effect\cite{Saitoh06,Kimura07} (ISHE) can convert the spin current generated by certain means, for instance spin pumping\cite{Tserkovnyak02,Zhang04}, into an electric signal. A particularly intriguing phenomenon that involves both SHE and ISHE is the spin Hall magnetoresistance\cite{Nakayama13,Avci15,Chen13,Althammer13,Vlietstra13,Hahn13,Vlietstra13_2,Isasa14,Vlietstra14,Marmion14} (SMR), in which a charge current in an NM causes a spin accumulation at the edge of the sample due to SHE, yielding a finite spin current at the interface to a ferromagnet. Through ISHE, the spin current gives an electromotive force along the original charge current, effectively changing the magnetoresistance of the NM.

The two major ingredients that determine SMR are the spin diffusion\cite{Zhang00} in the NM and the spin current at the NM/ferromagnet interface. The spin diffusion part has been addressed in detail by Chen {\it et al.} for the NM/ferromagnetic insulator (NM/FMI) bilayer, such as Pt/Y$_{3}$Fe$_{5}$O$_{12}$ (Pt/YIG), and FMI/NM/FMI trilayer\cite{Chen13}. This approach solves the spin diffusion equation in the presence of SHE and ISHE in a self-consistent manner, where the spin current at the NM/FMI interface serves as a boundary condition. However, the interface spin current remains an external parameter for which experimental or numerical input is needed\cite{Jia11,Burrowes12}. On the other hand, a quantum tunneling formalism has emerged recently as an inexpensive tool to calculate the interface spin current from various material properties such as the insulating gap of the FMI and the interface $s-d$ coupling\cite{Chen15}. The quantum tunneling theory also successfully explains\cite{Chen16} the reduced spin pumping spin current when an additional oxide layer is inserted between NM and FMI\cite{Du13}. It is then of fundamental importance to combine the spin diffusion approach with the quantum tunneling formalism for the interface spin current to give a complete theoretical description of the SMR, in particular to quantify how various material properties influence the SMR.

In this article we provide a minimal formalism that bridges the quantum tunneling formalism to the spin diffusion approach. We focus on the SMR in NM/FMI bilayer realized in Pt/YIG, and the NM/ferromagnetic metal (NM/FMM) bilayer realized in Pt/Co and Ta/Co\cite{Avci15}. The spin diffusion in the NM is assumed to be described by the same formalism of Chen {\it et al.}\cite{Chen13}, whereas the interface spin current is calculated from the quantum tunneling formalism\cite{Chen15,Chen16}. In the NM/FMM bilayer, we consider an FMM that has long spin diffusion length and a small thickness, such that the spin diffusion effect is negligible and the spin transport is predominately of quantum origin\cite{Chen15}. This is presumably adequate for the case of ultrathin Co films\cite{Bass07}, but not for materials with very short spin diffusion length such as permalloy\cite{Steenwyk97,Dubois99}. Within this formalism, the effect of material properties including spin diffusion length of the NM, interface $s-d$ coupling, insulating gap of the FMI, and the thickness of each layer can all be treated on equal footing. In particular, we reveal the signature of quantum interference in SMR in NM/FMM bilayer, and discuss the situation in which it can be observed. 


The structure of the article is arranged in the following manner. In Sec.~II, we detail the quantum tunneling formalism for the interface spin current in the NM/FMI bilayer, and how it is adopted into the spin diffusion approach that describes the NM. Section III generalizes this recipe to the NM/FMM bilayer, and discuss the observability of the predicted signature of quantum interference in SMR. Section IV gives the concluding remark.

\section{NM/FMI bilayer}

\subsection{Interface spin current}
\label{section:STT}

\begin{figure}[h!]
   \centering
  \includegraphics[clip=true,width=0.99\columnwidth]{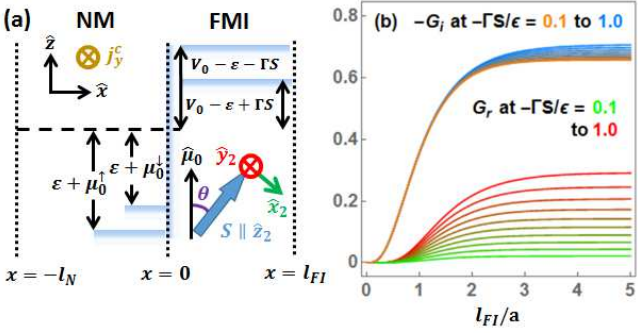}
   \caption{ (color online) (a) Schematics of the bilayer consists of an NM with thickness $l_{N}$ and an FMI with thickness $l_{FI}$. (b) The spin mixing conductance $G_{r,i}$ versus the FMI thickness $l_{FI}$, at different values of interface $s-d$ coupling strength $-\Gamma S/\epsilon_{F}$. The insulating gap strength is fixed at $(V_{0}-\epsilon_{F})/\epsilon_{F}=1.5$. The absolute units for $G_{r,i}$ is $e^{2}/\hbar a^{2}$ which is about $10^{14}\sim 10^{15}\Omega^{-1}$m$^{-2}$ depending on the Fermi wave length $a$. }
\label{fig:NMFMI_Gr_Gi}
\end{figure}

We start with the quantum tunneling formalism that calculates the interface spin current in the NM/FMI bilayer, which later serves as the boundary condition for the spin diffusion equation that determines SMR. The quantum tunneling formalism describes the NM/FMI bilayer shown in Fig. \ref{fig:NMFMI_Gr_Gi} (a) by the Hamiltonian
\begin{eqnarray}
H_{N} &=& \frac{p^{2}}{2m} - \mu_{x}^{\sigma} \;\;\; (-l_{N}\leq x<0) \: , \label{eq:Hamiltonian_NM} \\
H_{FI} &=& \frac{p^{2}}{2m} + V_{0} + \Gamma \boldsymbol{S}  \cdot \boldsymbol{\sigma} \;\;\; (0\leq x\leq l_{FI}) \: ,
\end{eqnarray}
where $\mu_{x}^{\sigma} = \pm \boldsymbol{\mu}_{x} \cdot \hat{\boldsymbol{z}}/2$ is the spin voltage of $\sigma = \{ \uparrow, \downarrow \}$ produced by an in-plane charge current $J_{y}^{c}{\hat{\bf y}}$, $\epsilon_{F}$ is the Fermi energy, $V_{0}-\epsilon_{F}$ is the insulating gap, and $\boldsymbol{S} = S(\sin \theta \cos \varphi, \sin \theta \sin \varphi, \cos \theta)$ is the magnetization. We choose $\Gamma < 0$ such that the magnetization has the tendency to align with the conduction electron spin $\boldsymbol{\sigma}$.
The wave function near the interface is
\begin{eqnarray}
   \psi_{N} &=& 
   (Ae^{ik_{0 \uparrow}x} + Be^{-ik_{0 \uparrow}x}) \left( \begin{array}{c} 1 \\ 0  \end{array} \right)
   +
   Ce^{-ik_{0 \downarrow}} \left( \begin{array}{c} 0 \\ 1 \end{array} \right) , \qquad 
   \label{eq:wave function_NM}\\
   \psi_{FI} &=& 
    (De^{q_{+}x} + Ee^{-q_{+}x}) \left( \begin{array}{c} e^{-i \varphi /2} \cos \frac{\theta}{2} \\ e^{i \varphi /2} \sin \frac{\theta}{2} \end{array} \right)
   \nonumber \\
    &&+
   (Fe^{q_{-}x} + Ge^{-q_{-}x}) \left( \begin{array}{c} -e^{-i \varphi /2} \sin \frac{\theta}{2} \\ e^{i \varphi /2} \cos \frac{\theta}{2} \end{array} \right),
   \label{eq:wave_equation_FI}
\end{eqnarray}
where $k_{0 \sigma} = \sqrt{2m(\epsilon_{F}+\mu_{0}^{\sigma})}/\hbar$ and $q_{\pm} = \sqrt{2m(V_{0} \pm \Gamma S - \epsilon_{F})}/\hbar$. The amplitudes $B \sim E$ are solved in terms of the incident amplitude $A$ by matching wave functions and their first derivative at the interface. The $x<-l_{N}$ and $x>l_{FI}$ regions are assumed to be vacuum or insulating oxides that correspond to infinite potentials such that the wave functions vanish there for simplicity. We identify the incident flux with $|A|^{2} = N_{F}|\boldsymbol{\mu}_{0}|/a^{3}$ where $N_{F}$ is the density of states per $a^{3}$ with $a=2\pi/k_{F}=h/\sqrt{2m\epsilon_{F}}$ the Fermi wave length.

The spin current inside the FMI at position $x$ is calculated from the evanescent wave function
\begin{eqnarray}
   \boldsymbol{j}_{x} = \frac{\hbar}{4im} \big[ \psi_{FI}^{*} \boldsymbol{\sigma} (\partial_{x} \psi_{FI}) - (\partial_{x} \psi_{FI}^{*}) \boldsymbol{\sigma} \psi_{FI} \big] \: .
\label{NMFMI_spin_current_general}
\end{eqnarray}
Angular momentum conservation\cite{Slonczewski96,Chen15} dictates that the interface spin current to be equal to the STT exerts on the magnetization
\begin{eqnarray}
&&\boldsymbol{j}_{0}-\boldsymbol{j}_{l_{FI}}=\boldsymbol{j}_{0}=\frac{\boldsymbol\tau}{a^{2}}
\nonumber \\ 
&&=\frac{\Gamma S N_{F}}{\hbar} \left[ G_{r} \hat{\boldsymbol{S}} \times \left( \hat{\boldsymbol{S}} \times \boldsymbol{\mu}_{0} \right) + G_{i} \hat{\boldsymbol{S}} \times \boldsymbol{\mu}_{0} \right],
\label{NMFMI_interface_spin_current}
\end{eqnarray}
which defines the field-like $G_{i}$ and dampling-like $G_{r}$ spin mixing conductance that in turn can be calculated from the interface spin current\cite{Chen15}
\begin{eqnarray}
\frac{\Gamma SN_{F}}{\hbar}G_{r}&=&\frac{2j_{0}^{x}\cos\varphi}{|{\boldsymbol \mu}_{0}|\sin 2\theta}+\frac{2j_{0}^{y}\sin\varphi}{|{\boldsymbol \mu}_{0}|\sin 2\theta}=-\frac{j_{0}^{z}}{|{\boldsymbol\mu}_{0}|\sin^{2}\theta}\;,
\nonumber \\
\frac{\Gamma SN_{F}}{\hbar}G_{i}&=&\frac{j_{0}^{x}\sin\varphi}{|{\boldsymbol \mu}_{0}|\sin \theta}-\frac{j_{0}^{y}\cos\varphi}{|{\boldsymbol \mu}_{0}|\sin \theta}\;.
\label{mixing_from_Js}
\end{eqnarray}
A straight forward calculation yields 
\begin{eqnarray}
   G_{r,i} &=& 
   \frac{-4}{a^{3} |\gamma_{\theta}|^{2}} \left( \frac{q_{+} \coth{q_{+}l_{FI}} - q_{-} \coth{q_{-}l_{FI}}}{q_{+}^{2} - q_{-}^{2}} \right)
   \nonumber \\ &&
   \times (\operatorname{Im,Re}) \; \Big( n_{\downarrow +}^{*} n_{\downarrow -} \Big) \: ,
\label{NMFMI_Gr_Gi}
\end{eqnarray}
where $\sigma_{x,y}$ is $x,y$ component of Pauli matrix, and
\begin{eqnarray}
n_{\sigma \pm} &=& \frac{k_{0 \sigma}}{(k_{0 \sigma} + i q_{\pm} \coth q_{\pm} l_{FI})}\;,
\nonumber \\
   \gamma_{\theta} &=& \frac{n_{\downarrow +}}{n_{\uparrow +}} \cos^{2} \frac{\theta}{2} + \frac{n_{\downarrow -}}{n_{\uparrow -}} \sin^{2} \frac{\theta}{2} \: .
\end{eqnarray}
Equation (\ref{NMFMI_Gr_Gi}) describes the spin mixing conductance in STT, as well as that in spin pumping since the Onsager relation\cite{Maekawa12} is satisfied in this approach\cite{Chen15}. Both $G_{r}$ and $G_{i}$ have very weak dependence (at most few percent) on the angle of magnetization $\theta$ through $\gamma_{\theta}$, which may be considered as higher order contributions\cite{Chen15}. In the numerical calculation below we set $\theta=0.3\pi$ without loss of generality.

Numerical results of the spin mixing conductance $G_{r,i}$ are shown in Fig.~\ref{fig:NMFMI_Gr_Gi}, plotted as a function the FMI thickness $l_{FI}$ and at different strength of the interface $s-d$ coupling $\Gamma S/\epsilon_{F}$. Both $G_{r}$ and $G_{i}$ increase with $l_{FI}$ initially and then saturate to a constant as expected, since they originate from the quantum tunneling of conduction electrons that only penetrate into the FMI over a very short distance. At a FMI thickness small compared to Fermi wave length $l_{FI}\ll a$, we found that $G_{r}\propto l_{FI}^{6}$ and $G_{i}\propto l_{FI}^{3}$, therefore the damping-like to field-like ratio is $|G_{r}/G_{i}|\ll 1$. In most of the parameter space, the torque is dominated by field-like component $|G_{r}/G_{i}|< 1$ throughout the whole range of $l_{FI}$. Only when the magnitude of $s-d$ coupling is large compared to the insulating gap $(V_{0}-\epsilon_{F})/\epsilon_{F}$ is the torque dominated by the damping-like component $|G_{r}/G_{i}|> 1$, consistent with that found previously\cite{Chen15} and also in accordance with the result from first principle calculation\cite{Jia11}. The magnitude of $G_{r,i}$ generally increases with the $s-d$ coupling, yet more dramatically for $G_{r}$. Note that $G_{r}$ and $G_{i}$ do not depend on the NM thickness in this quantum tunneling approach.

\begin{figure*}[ht]
   \centering
   \includegraphics[clip=true,width=1.99\columnwidth]{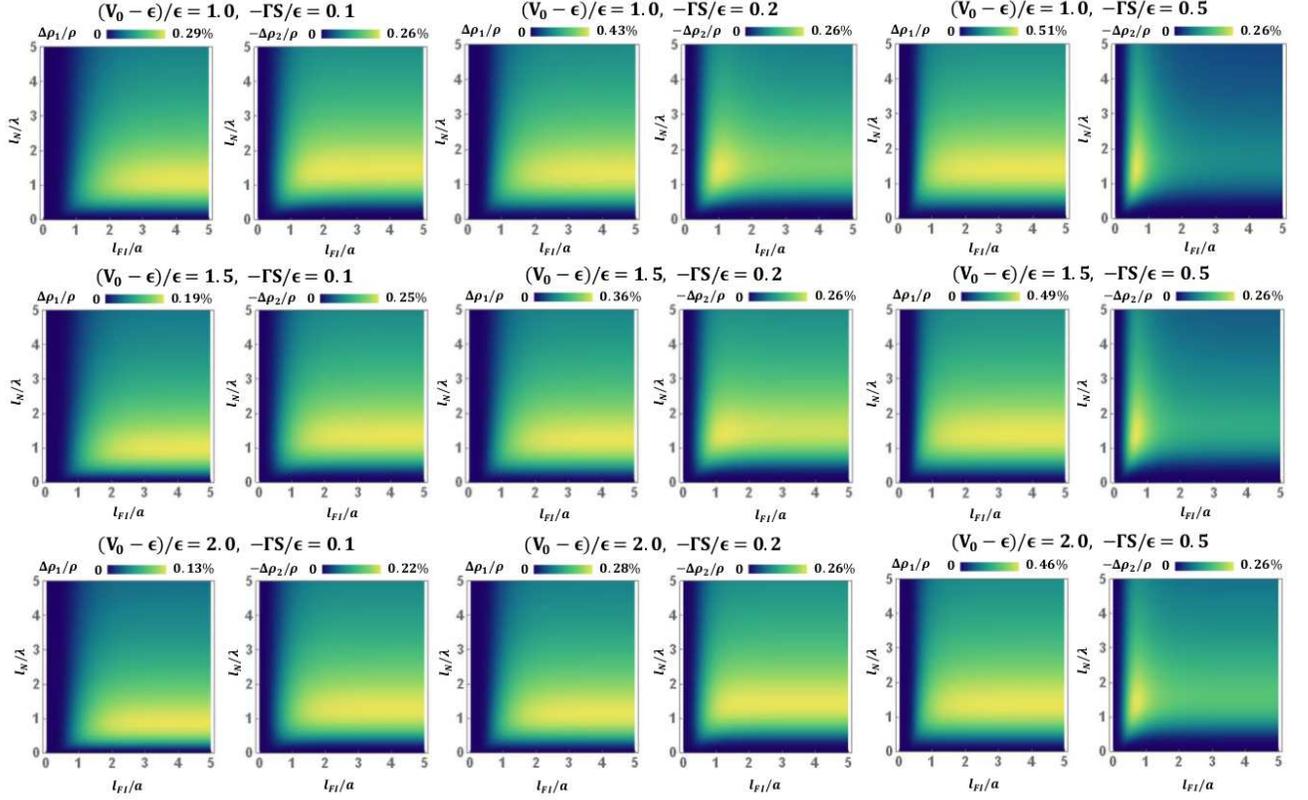}
   \caption{ (color online) The longitudinal $\Delta\rho_{1}/\rho$ and transverse $-\Delta\rho_{2}/\rho$ component of SMR in the NM/FMI bilayer, plotted against the FMI thickness in units of Fermi wave length $l_{FI}/a$ and NM thickness in units of the spin diffusion length $l_{N}/\lambda$, at various strength of $s-d$ coupling $-\Gamma S/\epsilon_{F}$ and the insulating gap $(V_{0}-\epsilon_{F})/\epsilon_{F}$. Note that the color scale of each plot is different. }
\label{fig:NMFMI_SMR}
\end{figure*}

\subsection{SMR}
\label{section:SMR}

We adopt the spin diffusion approach of Chen {\it et al.}\cite{Chen13} to address the effect of the interface spin current in Eq.~(\ref{NMFMI_interface_spin_current}) on SMR, which is briefly summarized below. The spin diffusion approach is based on the following assumptions for the spin transport in the NM: (1) The spin current in NM consists of two parts, one from the spatial gradient of spin voltage and the other the bare spin current caused directly by SHE, 
   \begin{eqnarray}
      \boldsymbol{j}_{x} = - \frac{\sigma_{c}}{4e^{2}} \partial_{x} \boldsymbol{\mu}_{x} + \frac{\theta_{\textrm{SH}} \sigma_{c} E_{y}}{2e} \hat{\boldsymbol{z}} \: ,
   \end{eqnarray}
where $\theta_{\textrm{SH}}$ is spin Hall angle, $\sigma_{c}$ is the conductivity of NM, $E_{y}$ is applied external electric in $y$ direction, and $-e$ is electron charge. (2) The spin voltage obeys the spin diffusion equation $\nabla^{2} \boldsymbol{\mu}_{x} = \boldsymbol{\mu}_{x} / \lambda^{2}$, where $\lambda$ is the spin diffusion length. (3) Spin current vanishes at the edge of NM ($x=-l_{N}$), which serves as one boundary condition. (4) The spin current at the NM/FMI interface is described by Eq.~(\ref{NMFMI_interface_spin_current}), which serves as another boundary condition. The self-consistent solution satisfying (1)$\sim$(4) is\cite{Chen13} 
\begin{eqnarray}
   \frac{\boldsymbol{j}_{x} \cdot \hat{\boldsymbol{x}}}{j_{\textrm{SH}}}
   &=&
   \beta_{x}\sin\theta
   \Big[ 
   \cos \theta \cos \varphi 
   \operatorname{Re} \big( \widetilde{G} \big) 
   +
   \sin \varphi
   \operatorname{Im} \big( \widetilde{G} \big)
   \Big] ,
   \nonumber \\
   \frac{\boldsymbol{j}_{x} \cdot \hat{\boldsymbol{y}}}{j_{\textrm{SH}}}
   &=&
   \beta_{x}\sin\theta
   \Big[ 
   \cos \theta \sin \varphi 
   \operatorname{Re} \big( \widetilde{G} \big) 
   -
   \cos \varphi
   \operatorname{Im} \big( \widetilde{G} \big)
   \Big] ,
   \nonumber \\
   \frac{\boldsymbol{j}_{x} \cdot \hat{\boldsymbol{z}}}{j_{\textrm{SH}}}
   &=&
   1 - \frac{\cosh (\frac{2x+l_{N}}{2 \lambda})}{\cosh (\frac{l_{N}}{2 \lambda})}
   -
   \beta_{x} \sin^{2} \theta
   \operatorname{Re} \big( \widetilde{G} \big) ,
\label{NMFMI_js}
\end{eqnarray}
where
\begin{eqnarray}
   \beta_{x} &=& \frac{\sinh (\frac{x+l_{N}}{\lambda})}{\sinh (\frac{l_{N}}{\lambda})} \tanh (\frac{l_{N}}{2 \lambda})\;,
\nonumber   \\
   \widetilde{G} &=& \frac{\alpha G_{c}}{1 - \alpha G_{c}\coth (\frac{l_{N}}{\lambda})} \: ,
\nonumber \\
\alpha &=& \frac{4 \Gamma S N_{F}e^{2} \lambda}{ \hbar \sigma_{c}}\;,
\label{NMFMI_Gtilde}
\end{eqnarray} 
and $j_{\textrm{SH}} = \theta_{\textrm{SH}} \sigma_{c} E_{y} / 2e$ is the bare spin current. Here $\alpha <0$ is a negative parameter (because we assume the interface $s-d$ coupling $\Gamma<0$) that bridges our tunneling formalism to the spin diffusion equation, and $G_{c}=G_{r}+iG_{i}$ is the complex spin mixing conductance.


Through ISHE, the spin currents in Eq.~(\ref{NMFMI_js}) is converted back to a charge current in the longitudinal (along $\hat{\boldsymbol{y}}$) and transverse (along $\hat{\boldsymbol{z}}$) direction
\begin{eqnarray}
   \Delta j_{\textrm{long}}^{c}(x) &=& -2 e \theta_{\textrm{SH}} \Big( \boldsymbol{j}_{x} - \frac{\theta_{\textrm{SH}} \sigma_{c} E_{y}}{2e} \hat{\boldsymbol{z}} \Big) \cdot \hat{\boldsymbol{z}},
   \\
   \Delta j_{\textrm{trans}}^{c}(x) &=& 2 e \theta_{\textrm{SH}} \Big( \boldsymbol{j}_{x} - \frac{\theta_{\textrm{SH}} \sigma_{c} E_{y}}{2e} \hat{\boldsymbol{z}} \Big) \cdot \hat{\boldsymbol{y}}.
\end{eqnarray}
The conductivity averaged over the NM layer then follows
\begin{eqnarray}
   \sigma_{\textrm{long}} &=& \sigma + \frac{1}{l_{N} E_{y}} \int_{-l_{N}}^{0} dx \: \Delta j_{\textrm{long}}^{c}(x) ,
   \\
   \sigma_{\textrm{trans}} &=& \frac{1}{l_{N} E_{y}} \int_{-l_{N}}^{0} dx \: \Delta j_{\textrm{trans}}^{c}(x) .
\end{eqnarray}
Using $\theta_{\textrm{SH}}^{2} \sim 0.01 \ll 1$, the longitudinal and transverse component of SMR read
\begin{eqnarray}
   \rho_{\textrm{long}} &=& \sigma_{\textrm{long}}^{-1} \approx \rho + \Delta \rho_{0} + \sin^{2} \theta \: \Delta \rho_{1} ,
\nonumber \\
   \rho_{\textrm{trans}} &=& -\sigma_{\textrm{trans}} / \sigma_{\textrm{long}}^{2}  
\nonumber \\
&\approx& \cos \theta \sin \theta \sin \varphi \: \Delta \rho_{1} - \sin \theta \cos \varphi \: \Delta \rho_{2} ,
   \label{eq:rho1} 
\end{eqnarray}
where
\begin{eqnarray}
   \Delta \rho_{0} / \rho &=& - \theta_{\textrm{SH}}^{2} \frac{2 \lambda}{l_{N}} \tanh\left(\frac{l_{N}}{2\lambda}\right) ,
\nonumber \\
   \Delta \rho_{1} / \rho &=& - \theta_{\textrm{SH}}^{2} \frac{\lambda}{l_{N}}\tanh^{2}\left(\frac{l_{N}}{2\lambda}\right) \operatorname{Re} \big( \widetilde{G} \big),
 \nonumber \\
   \Delta \rho_{2} / \rho &=&  \theta_{\textrm{SH}}^{2} \frac{\lambda}{l_{N}} \tanh^{2}\left(\frac{l_{N}}{2\lambda}\right) \operatorname{Im} \big( \widetilde{G} \big). 
\label{eq:rho5}
\end{eqnarray}
Clearly the FMI thickness $l_{FI}$ affects $\Delta\rho_{1}$ and $\Delta\rho_{2}$ only through $\widetilde{G} = \widetilde{G} (l_{FI})$.

To perform numerical calculation of Eq.~(\ref{eq:rho5}), we make the following assumption on the parameter $\alpha$ in Eq.~(\ref{NMFMI_Gtilde}) that connects the quantum tunneling formalism with the spin diffusion equation. Firstly, $\alpha$ contains the density of state per $a^{3}$ at the Fermi surface, which is assumed to be the inverse of Fermi energy $N_{F}=1/\epsilon_{F}$. The combined parameter $\Gamma SN_{F}=\Gamma S/\epsilon_{F}$ therefore represents the strength of $s-d$ coupling. Other parameters that influence $\alpha$ are the spin diffusion length assumed to be $\lambda\approx 10$nm, the conductivity of the NM film taken to be $\sigma_{c}\approx 5\times 10^{6}\Omega^{-1}$m$^{-1}$, and Fermi wave length assumed to be roughly equal to the lattice constant $a\approx 0.4$nm, all of which are the typical values for commonly used materials such as Pt. These lead to the dimensionless parameter $\alpha G_{c}\approx 10\times\left(\Gamma S/\epsilon_{F}\right)\times\left(G_{c}/(e^{2}/\hbar a^{2})\right)$ in Eq.~(\ref{NMFMI_Gtilde}) being expressed in terms of the relative strength of $s-d$ coupling and the spin mixing conductance divided by its unit. In what follows, we examine the effect of FMI thickness, NM thickness, insulating gap, and interface $s-d$ coupling on SMR. On the contrary, the spin Hall angle, spin diffusion length, and conductivity are treated as constants, although in reality they may also depend on the layer thickness or on each other in such thin films\cite{Sagasta16}.


The numerical result of SMR is shown in Fig.~\ref{fig:NMFMI_SMR}, plotted as a function of the FMI thickness $l_{FI}$ and NM thickness $l_{N}$ at several values of insulating gap $(V_{0}-\epsilon_{F})/\epsilon_{F}$ and $s-d$ coupling $\Gamma S/\epsilon_{F}$. As a function of the FMI thickness $l_{FI}$, both the longitudinal $\Delta\rho_{1}/\rho$ and the transverse $\Delta\rho_{2}/\rho$ component initially increase and then saturate at around $l_{FI}/a\sim 2$, which is expected since conduction electrons only tunnel into the FMI over a short depth, so the interface spin current saturates once the FMI is thicker than this tunneling depth. The insulating gap $(V_{0}-\epsilon_{F})/\epsilon_{F}$ obviously affects the tunneling depth, and is particularly influential on the magnitude of longitudinal $\Delta\rho_{1}/\rho$, as can be seen by comparing plots with different $(V_{0}-\epsilon_{F})/\epsilon_{F}$ in Fig.~\ref{fig:NMFMI_SMR}. The magnitude of $\Delta\rho_{1}/\rho$ also generally increases with the $s-d$ coupling $\Gamma S/\epsilon_{F}$, while the transverse component $\Delta\rho_{2}/\rho$ at large $\Gamma S/\epsilon_{F}$ displays a nonmonotonic dependence on the FMI thickness. On the other hand, as a function of NM thickness $l_{N}$, both SMR components increase and peak at around $l_{N}/\lambda\sim 1$ and then decrease monotonically for large $l_{N}$. This can be understood because both $\Delta\rho_{1}$ and $\Delta\rho_{2}$ are interface effects that become less significant compared to bulk resistivity $\rho$ when NM thickness increases, and the spin voltage is known to be maximal when the NM thickness is comparable to the spin diffusion length\cite{Zhang00} $l_{N}/\lambda\sim 1$.



\section{NM/FMM BILAYER}

\subsection{Interface spin current and SMR}

\begin{figure}[h!]
   \centering
  \includegraphics[clip=true,width=0.99\columnwidth]{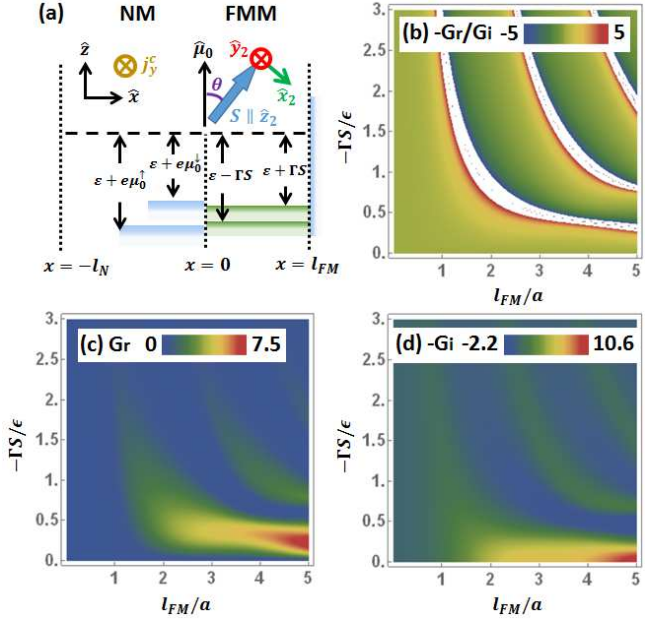}
   \caption{ (color online) (a) Schematics of an NM/FMM bilayer with finite thickness. (b) The ratio of spin mixing conductance (c) $G_{r}$ and (d) $-G_{i}$ in this system, plotted against the thickness $l_{FM}$ of the FMM and $s-d$ coupling $-\Gamma S/\epsilon_{F}$, in units of $e^{2}\hbar/a^{2}$ where $a$ is the Fermi wave length. }
\label{fig:NMFMMfinite_Gr_Gi}
\end{figure}

We proceed to address the SMR in the NM/FMM bilayer, with the assumption that the FMM film is much thinner than its spin diffusion length $l_{FM}\ll\lambda$ such that quantum tunneling is the dominant mechanism for spin transport in the FMM, while the spin diffusion inside the FMM can be ignored. The calculation of the spin current at the NM/FMM interface starts with the model schematically shown in Fig. \ref{fig:NMFMMfinite_Gr_Gi} (a). The NM and FMM occupy $-l_{N} \leq x < 0$ and $0 \leq x \leq l_{FM}$, respectively. The NM region is described by Eqs. (\ref{eq:Hamiltonian_NM}) and (\ref{eq:wave function_NM}), while the FMM layer is described by $H_{FM} = p^{2}/2m +\Gamma{\bf S}\cdot{\boldsymbol\sigma}$ and the wave function 
\begin{eqnarray}
   \psi_{FM} &=& 
   (De^{ik_{+}x} + F^{-ik_{+}x}) \left( \begin{array}{c} e^{-i \varphi /2} \cos \frac{\theta}{2} \\ e^{i \varphi /2} \sin \frac{\theta}{2} \end{array} \right)
   \nonumber \\ &&
    + (E e^{ik_{-}x} + Ge^{-ik_{-}x}) \left( \begin{array}{c} -e^{-i \varphi /2} \sin \frac{\theta}{2} \\ e^{i \varphi /2} \cos \frac{\theta}{2} \end{array} \right)
    \: , \qquad
\end{eqnarray}
where $k_{\pm} = \sqrt{2m(\epsilon_{F} \mp \Gamma S)}/ \hbar$. The wave functions outside of the bilayer in $x>l_{FM}$ and $x<-l_{N}$ are assumed to vanish for simplicity. The coefficients $A \sim I$ are again determined by matching wave functions and their first derivative at the interface. The interface spin current and the spin mixing conductance are calculated from Eqs.~(\ref{NMFMI_spin_current_general}) to (\ref{mixing_from_Js}), with replacing $\psi_{FI}$ to $\psi_{FM}$ and $l_{FI}$ to $l_{FM}$, resulting in
\begin{eqnarray}
   G_{r,i} &=& \frac{1}{a^{3} |\gamma_{\theta}'|^{2}} (\operatorname{Im,Re}) \Bigg[ Z_{\downarrow - +}^{*} Z_{\downarrow + +}
\nonumber \\
&&\times\Big( 
   u_{+-} - u_{++} - u_{--} + u_{-+}
   \Big) \Bigg] \: ,
\label{eq:spin_mixing_conductance_NM/FMM}
\end{eqnarray}
where
\begin{eqnarray}
    u_{\alpha \beta} &=& i \frac{e^{i(\alpha k_{+} + \beta k_{-})l_{FM}}}{\alpha k_{+} + \beta k_{-}} \quad , \quad 
   W_{\sigma \alpha \beta} = \frac{k_{0 \sigma} + \beta k_{\alpha}}{2 k_{0 \sigma}} \quad , \nonumber \\
   Z_{\sigma \alpha \beta} &=& W_{\sigma \alpha \beta} e^{-ik_{\alpha}l_{FM}} - W_{\sigma \alpha \overline{\beta}} e^{ik_{\alpha}l_{FM}} \quad , \nonumber \\
   \gamma'_{\theta} &=& Z_{\uparrow ++} Z_{\downarrow -+} \cos^{2} \frac{\theta}{2} + Z_{\downarrow ++} Z_{\uparrow -+} \sin^{2} \frac{\theta}{2} \: ,
\label{eq:FMM_coefficients_2}
\end{eqnarray}
with $\overline{\beta}=-\beta$. Apart from a change in magnitude, the pattern of the spin mixing conductance as a function of $s-d$ coupling and the FMM thickness shown in Fig.~\ref{fig:NMFMMfinite_Gr_Gi} is almost indistinguishable from that reported in Fig.~2 of Ref.~\onlinecite{Chen15}, which shows clear signals of quantum interference with respect to both $s-d$ coupling and FMM thickness. This similarity is expected, since the only difference between the formalism here and in Ref.~\onlinecite{Chen15} is the insulating gap $V_{0}-\epsilon_{F}$ of the substrate or vacuum in the $x>l_{FM}$ region in Fig.~\ref{fig:NMFMMfinite_Gr_Gi} (a), which is assumed to be infinite here for simplicity but finite in Ref.~\onlinecite{Chen15}. The insulating gap is spin degenerate and essentially does not influence the spin transport.

\begin{figure}[h!]
   \centering
  \includegraphics[clip=true,width=0.99\columnwidth]{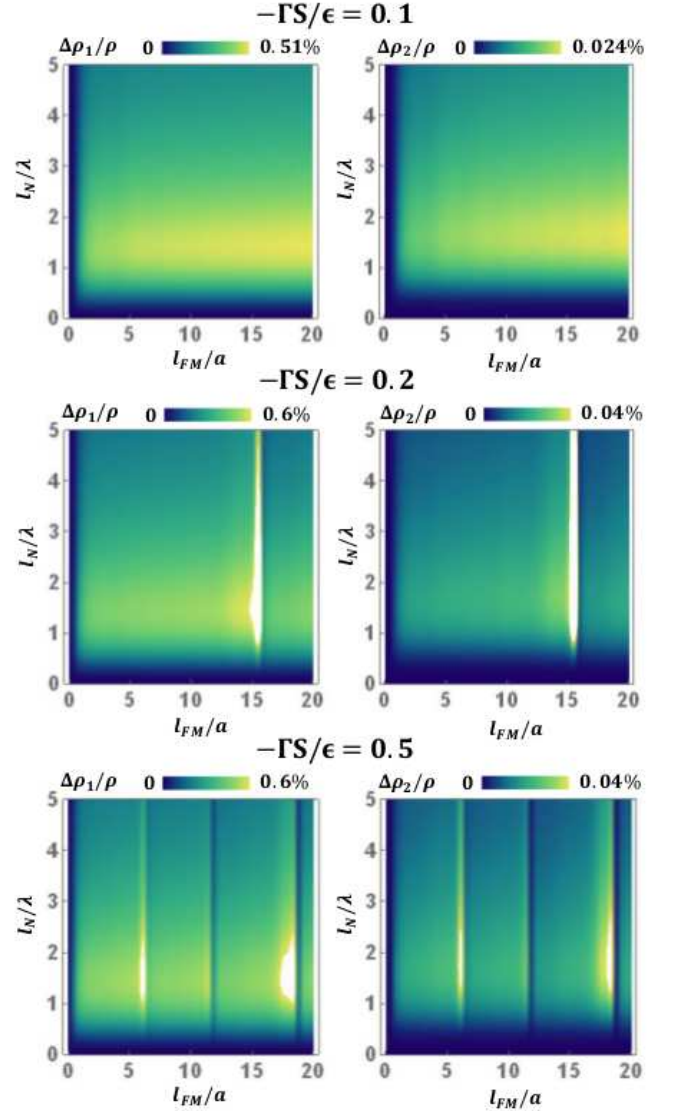}
   \caption{ (color online) The longitudinal $\Delta\rho_{1}/\rho$ and transverse $\Delta\rho_{2}/\rho$ component of SMR in the NM/FMM bilayer, plotted against the FMM thickness in units of Fermi wave length $l_{FM}/a$ and NM thickness in units of the spin diffusion length $l_{N}/\lambda$, at various strength of $s-d$ coupling $-\Gamma S/\epsilon_{F}$. }
\label{fig:NMFMM_SMR}
\end{figure}

To get SMR, we use Eq. (\ref{eq:rho1}) $\sim$ (\ref{eq:rho5}) while taking the $G_{c} = G_{r}+ iG_{i}$ obtained from Eq.~(\ref{eq:spin_mixing_conductance_NM/FMM}). The results for the longitudinal $\Delta\rho_{1}/\rho$ and transverse $\Delta\rho_{2}/\rho$ component of SMR as functions of FMM thickness $l_{FM}$ and NM thickness $l_{N}$ are shown in Fig.~\ref{fig:NMFMM_SMR}, for several values of $s-d$ coupling $\Gamma S/\epsilon_{F}$. As a function of NM thickness, both components reach a maximal at around the spin diffusion length $l_{N}/\lambda\sim 1$ and then decrease monotonically, similar to that reported in Fig.~\ref{fig:NMFMI_SMR} for NM/FMI bilayer and is due to the spin diffusion effect explained in Sec.~\ref{section:SMR}. On the other hand, as a function of FMM thickness, both components show clear modulations with an average periodicity that decreases with increasing $s-d$ coupling, a trend similar to that of $G_{r}$ and $G_{i}$ shown in Fig.~\ref{fig:NMFMMfinite_Gr_Gi} and is attributed to the quantum interference of spin transport. Intuitively, a larger $s-d$ coupling renders a faster precession of conduction electron spin when it travels inside the FMM, hence more modulations appear for a given FMM thickness. The transverse component of SMR is found to be generally one order of magnitude smaller than the longitudinal component.

\subsection{To observe the predicted oscillation in SMR}

The experimental detection of the oscillation of SMR with respect to FMM thickness $l_{FM}$ shown in Fig.~\ref{fig:NMFMM_SMR} would be a direct proof of our approach. In a typical NM/FMI set up, however, there are other sources that contribute to the total resistance measured in experiments, therefore it is important to investigate whether there is a situation in which the predicted oscillation of SMR can manifest. To explore this possibility, we use a three-resistor model to characterize the total longitudinal resistance\cite{Avci15}, which contains the resistor that represents the NM layer ($N$), the FMM layer ($F$), and the interface layer ($I$) connected in parallel, each denoted by $R_{i}=R_{i}^{0}+\delta R_{i}$ with $i=\left\{N,I,F\right\}$. Here $R_{i}^{0}$ is the contribution to the longitudinal resistance in layer $i$ that does not depend on the angle of the magnetization, and $\delta R_{i}$ is the part that depends on the angle which is generally much smaller $\delta R_{i}\ll R_{i}^{0}$. Expanding the total longitudinal resistance to leading order in $\delta R_{i}$ yields 
\begin{eqnarray}
R_{tot}
&\approx& R_{tot}^{0}+\left(\frac{R_{I}^{0}R_{F}^{0}}{B}\right)^{2}\delta R_{N}
\nonumber \\
&&+\left(\frac{R_{N}^{0}R_{F}^{0}}{B}\right)^{2}\delta R_{I}
+\left(\frac{R_{N}^{0}R_{I}^{0}}{B}\right)^{2}\delta R_{F}\;,
\nonumber \\
R_{tot}^{0}&=&\frac{R_{N}^{0}R_{I}^{0}R_{F}^{0}}{B}\;,
\nonumber \\
B&=&R_{N}^{0}R_{I}^{0}+R_{I}^{0}R_{F}^{0}+R_{N}^{0}R_{F}^{0}\;.
\end{eqnarray}  
Each resistance is assumed to satisfy the usual relation to the sample size $\left\{R_{i}^{0},\delta R_{i}\right\}=\left\{\rho_{i}^{0},\delta \rho_{i}\right\}\times L/l_{i}\;h$, where $L$ and $h$ are the length and the width of the sample, respectively, $\rho_{i}^{0}$ and $\delta\rho_{i}$ are the corresponding resistivity, and $l_{i}$ is the thickness of layer $i$. The thickness of the interface $l_{I}$ is assumed to be intrinsically constant, in contrast to $l_{N}$ and $l_{F}$ that can be varied experimentally\cite{Avci15}. The percentage change of the total resistance due to the angle of the magnetization is 
\begin{eqnarray}
\frac{R_{tot}-R_{tot}^{0}}{R_{tot}^{0}}&\approx&\left(\frac{\rho_{I}^{0}\rho_{F}^{0}}{l_{I}l_{F}C}\right)\frac{\delta\rho_{N}}{\rho_{N}^{0}}
+\left(\frac{\rho_{N}^{0}\rho_{F}^{0}}{l_{N}l_{F}C}\right)\frac{\delta\rho_{I}}{\rho_{I}^{0}}
\nonumber \\
&&+\left(\frac{\rho_{N}^{0}\rho_{I}^{0}}{l_{N}l_{I}C}\right)\frac{\delta\rho_{F}}{\rho_{F}^{0}}\;,
\nonumber \\
C&=&\frac{\rho_{N}^{0}\rho_{I}^{0}}{l_{N}l_{I}}+\frac{\rho_{N}^{0}\rho_{F}^{0}}{l_{N}l_{F}}+\frac{\rho_{F}^{0}\rho_{I}^{0}}{l_{F}l_{I}}\;.
\label{Rtot_percentage}
\end{eqnarray} 
Note that the $\rho_{i}^{0}\rho_{j}^{0}/l_{i}l_{j}C$ factors are monotonic functions of the layer thickness $\left\{l_{N},l_{I},l_{F}\right\}$, and are independent from the angle of the magnetization.

The contribution to the angular dependent part of $R_{F}$ comes from the anisotropic magnetoresistance (AMR) which takes the form\cite{McGuire75,Zhang15} $\delta\rho_{F}\propto({\boldsymbol j}^{c}\cdot{\hat{\bf m}})^{2}\propto (m^{y})^{2}$ since the in-plane charge current ${\boldsymbol j}^{c}$ runs along ${\hat{\bf y}}$ as shown in Fig.~\ref{fig:NMFMMfinite_Gr_Gi} (a), and we denote ${\hat{\bf m}}={\bf S}/S=(\sin\theta\cos\varphi,\sin\theta\sin\varphi,\cos\theta)$ as the unit vector along the direction of the magnetization. In addition, Zhang {\it et al.}\cite{Zhang15} showed that the interface resistance has a quadratic dependence on both $m^{y}$ and $m^{z}$, a result of surface spin-orbit scattering. On the other hand, the SMR in the NM has the angular dependence\cite{Chen13} described by Eq.~(\ref{eq:rho1}). These considerations lead to the parametrization of resistivity by
\begin{eqnarray}
&&\rho_{F}^{0}+\delta\rho_{F}=\rho_{F}^{0}+\Delta\rho_{F}^{b}(m^{y})^{2}\;,
\nonumber \\
&&\rho_{I}^{0}+\delta\rho_{I}=\rho_{I}^{0}+\Delta\rho_{I,y}^{s}(m^{y})^{2}+\Delta\rho_{I,z}^{s}(m^{z})^{2}\;,
\nonumber \\
&&\rho_{N}^{0}+\delta\rho_{N}=\left(\rho+\Delta\rho_{0}\right)+\Delta\rho_{1}\left[(m^{x})^{2}+(m^{y})^{2}\right]\;.
\nonumber \\
\label{rho_NIF}
\end{eqnarray}
Combinig this with Eq.~(\ref{Rtot_percentage}) motivates us to propose the following experiment that should isolate the effect of longitudinal SMR represented by $\delta\rho_{N}$. From Eq.~(\ref{rho_NIF}), we see that $\delta\rho_{F}$ and $\delta\rho_{I}$ vanish if the magnetization does not have an in-plane component, i.e., $m^{y}=m^{z}=0$, while $\delta\rho_{N}$ remains finite as long as the out-of-plane component is nonzero $m^{x}\neq 0$. Thus we propose to fix the magnetization of the FMM film to be out-of-plane $m^{x}\neq 0$, in which case the percentage change of total longitudinal resistance as a function of FMM thickness takes the form 
\begin{eqnarray}
\frac{R_{tot}-R_{tot}^{0}}{R_{tot}^{0}}\approx\frac{l_{1}}{l_{F}+l_{1}+l_{2}}\times\frac{\Delta\rho_{1}}{\rho+\Delta\rho_{0}}(m^{x})^{2}\;,&&
\nonumber \\
({\rm for}\;\;m^{y}=m^{z}=0)&&
\label{Rtot_SySz0}
\end{eqnarray}
where $\rho$, $\Delta\rho_{0}$, and $\Delta\rho_{1}$ are those in Eqs.~(\ref{eq:rho1}) and (\ref{eq:rho5}), $l_{1}=l_{N}\rho_{F}^{0}/\rho_{N}^{0}$ and $l_{2}=l_{I}\rho_{F}^{0}/\rho_{I}^{0}$ are two length scales that can be treated as fitting parameters in experiments. Equation (\ref{Rtot_SySz0}) indicates that, for the case of only out-of-plane magnetization, the percentage change of magnetoresistance decays with the FMM thickness $l_{F}$ due to the $l_{1}/(l_{F}+l_{1}+l_{2})$ factor, but also oscillates with $l_{F}$ due to the $\Delta\rho_{1}/(\rho+\Delta\rho_{0})\approx\Delta\rho_{1}/\rho$ factor as quantified in Eq.~(\ref{eq:rho5}) and shown in Fig.~\ref{fig:NMFMM_SMR}. Thus varying FMM thickness while keeping its magnetization out-of-plane may be a proper set up to observe the predicted oscillation of longitudinal SMR, provided the FMM thickness remains thinner than its spin relaxation length $l_{F}\ll\lambda$. Finally, we remark that the convention of labeling coordinate in SMR or STT experiments is that the charge current is defined to be along ${\hat x}$ and the direction normal to the film is along ${\hat z}$. Therefore the coordinate in our tunneling formalism $(x,y,z)$ corresponds to $(z,x,y)$ in the experimental convention.


\section{Conclusion}

In summary, the quantum tunneling formalism for the interface spin current is incorporated into the spin diffusion approach to study the effect of various material properties on SMR, in particular the effect of layer thickness, insulating gap, and interface $s-d$ coupling. The advantage of combining the quantum and diffusive approach is that the effects of all these material properties can be treated on equal footing. For the NM/FMI case, we reveal an SMR that saturates at large FMI thickness since the conduction electrons only tunnels into the FMI over a short distance, whereas the longitudinal and transverse SMR display different dependence on the insulating gap and interface $s-d$ coupling. For the NM/FMM case, we predict that SMR may display a pattern of oscillation as increasing FMM thickness due to quantum interference, and propose an experiment to observe it by using fixed out-of-plane magnetization to isolate SMR from other contributions. We anticipate that our minimal model that combines the quantum and diffusive approach may be used to guide the search for suitable materials that optimize the SMR, and help to predict novel spin transport effects in ultrathin heterostructures in which quantum effects shall not be overlooked.

The authors acknowledge the fruitful discussions with P. Gambardella, F. Casanova, and J. Mendil. W. C. and M. S. are grateful for the financial support
through a research grant of the Swiss National Science Foundation.

\end{document}